\documentclass[10pt,tightenlines,
    onecolumn,
    preprintnumbers,
    amsmath,
    amssymb,
    prd,
    showkeys,
    showpacs
]{revtex4}
\usepackage{color}
\usepackage{graphicx}%
\usepackage{amsmath}
\usepackage[normalem,normalbf]{ulem}
\usepackage{amssymb}
\usepackage{bm}
\usepackage{enumerate}
\newcommand{\nn}{\nonumber}

\begin{document}
%-------------------------%-------------------------------------
\title{Symmetric ordering effect on Casimir energy in $\kappa-$Minkowski spacetime}
%------------------------------------------------------------

\author{Hyeong-Chan Kim$^1$}
\email{hckim@phya.yonsei.ac.kr}
%------------------------------------------------------------
\author{Chaiho Rim$^2$}
\email{rim@chonbuk.ac.kr}
%-------------------------------------------------
\author{Jae Hyung Yee$^1$}%
\email{jhyee@yonsei.ac.kr}
%-------------------------------------------------------------
\affiliation{$^1$ Department of Physics, Yonsei University,
Seoul 120-749, Republic of Korea\\
and\\
%}%
%-------------------------------------------------------------
%\affiliation{
$^2$ Department of Physics and Research Institute
of Physics and Chemistry, Chonbuk National University,
Jeonju 561-756, Korea.
}%
%\date{\today}%
%-------------------------------------------------------
\bigskip
%-------------------------------------------------------------
\begin{abstract}
%----------------------------------------------------------
We present the Casimir energy of spherical shell,
for the symmetrically deformed scalar field in $\kappa$-Minkowski space-time,
satisfying Dirichlet boundary condition. 
The Casimir energy shows the particle anti-particle 
symmetry contrary to the asymmetrically deformed case.
In addition, the deformation effect 
starts from $O(1/\kappa)$ term unlike in the parallel plates.

%----------------------------------------------
\end{abstract}
%------------------------------------------------------
%--------------------------------------------------------
%------------------------------------------------------------
\maketitle
%%%%%%%%%%%%%%%%%%%%%%%%%%%%%%%%%%%%%%%%%%%%%%%
\section{Introduction}
%%%%%%%%%%%%%%%%%%%%%%%%%%%%%%%%%%%%%%%%%%%%%%%

The Casimir energy is dependent on the geometry
as shown in~\cite{casimir,boyer}:
The Casimir force is attractive between parallel plates
but repulsive in a sphere.
When the spacetime is $\kappa$-deformed~\cite{ncst},
the Poincar\'{e} algebra is deformed~\cite{kappaP}
so that the energy-momentum relation from the 
Einstein special relativity is deformed. 
In this case, 
the Casimir energy provides a useful information 
about the vacuum structure of the theory 
as shown in~\cite{KRY-asym}.
It turns out that the $\kappa$-deformed Poincar\'{e} algebra 
has many different versions, 
which originate from 
the ordering of the space and time coordinates. 
In our previous paper in \cite{KRY-asym}, 
the Casimir energy for the so-called asymmetric ordering 
case is investigated,
where the vacuum is shown to break 
the particle and anti-particle symmetry.
Thus, one needs to investigate if there exists a case 
where the vacuum respects the particle and anti-particle 
symmetry. One obvious choice is the symmetric ordering
case, where particle and antiparticle dispersion relation 
is symmetric. 
In this brief report, we evaluate 
the Casimir energy for the particle and anti-particle 
contribution for the symmetric ordering case.
In section \ref{sec:2}, a
brief summary is given how to evaluate the Casimir energy.
In section \ref{sec:3},
the Casimir energy is given with an appropriate measure,
and summary and discussion is given in  section \ref{sec:4}.

%%%%%%%%%%%%%%%%%%%%%%%%%%%%%%%%%%%%%%%%%%%%%%%
\section{Casimir energy of a spherical shell} \label{sec:2}
%%%%%%%%%%%%%%%%%%%%%%%%%%%%%%%%%%%%%%%%%%%%%%%
The Casimir energy is the zero point vacuum energy of
massless scalar fields, which in momentum space is given by
$E_c^+ =  \frac 12 \int_{\bf p}
  \, \hbar \omega_{\bf p} $
where $ \int_{\bf p}$ denotes the momentum integration with
$\kappa$-Poincar\'{e} invariant measure $\int \frac{d^3 \bf
p}{(2\pi)^3} e^{{\alpha \omega_{\bf p}}/{\kappa } }$ with
$\alpha=3/2$, and
$\omega_{\bf p}$ is the positive mode (particle) dispersion relation for the symmetric ordering case
\begin{eqnarray}
\label{omega}
\omega_{\bf p}&=&2 \kappa \ln \left(\frac{|\bf p|}{2\kappa}
    +\sqrt{1+\frac{{\bf p}^2}{4\kappa ^2}} \,\right)
    =-2 \kappa \ln  \left(-\frac{|\bf p|}{2\kappa}
    +\sqrt{1+\frac{{\bf p}^2}{4\kappa ^2}} \,\right)\,.
\end{eqnarray}
The negative mode (anti-particle) contribution $E_c^-$
is obtained using the same $\omega_{\bf p}$ 
but with the measure changed $\int \frac{d^3 \bf
p}{(2\pi)^3} e^{{-\alpha \omega_{\bf p}}/{\kappa } }$.
We note 
that $E_c^-$ can be formally obtained 
if one changes the sign
of $\kappa$ to $-\kappa$.  
Following the prescription given in Ref.~\cite{KRY-asym}, 
we can put the Casimir energy in the form 
\begin{equation}
\label{eq:Casimir-largeorder}
E_c(a) = \sum_{n=0}^\infty \Big( {\cal E}_n (a)
- {\cal E}_n ( \eta R) \Big)
\end{equation}
where $a$ represents the radius of the sphere,
$\eta R$ is radius of a large sphere 
introduced to regularize the mode with $0<\eta<1$
and 
\begin{eqnarray}
\label{En}
{\cal E}_n (r)  &=&  - \sum_l  \frac{\kappa}{\pi\nu^{2n-1}}
\Re\int_0^\infty dy\,
i  \Big(e^{- i \sigma \nu ye^{-i \phi}}\, g(r, i \nu y e^{-i \phi})
)\Big)\frac{d}{dy}
        q_n(y e^{-i \phi})
\end{eqnarray}
where the limit ${R \to \infty, \sigma \to 0, \phi \to 0}$ is
assumed at the end. 
$g(r,z)\equiv -e^{{\alpha \omega(z, r)}\, 
/{\kappa } }\, {\omega(z, r)}/{\kappa}\,$ 
and $q_n(y)$ corresponds to the series expansion of the 
large order Bessel function 
whose explicit form for $n=0,1,2$ is given in~\cite{KRY-asym}. 
The merit of this decomposition is that
${\cal E}_0 (a) - {\cal E}_0( \eta R) =0$,
and all other components are finite,
and allow the systematic expansion
in $1/(\kappa a)$.  
${\cal E}_n (r) $ is conveniently put as,
after the integration by part
\begin{equation}
\label{En-analytic}
{\cal E}_n (r) = \frac1r \sum_l  \frac{B_n (\nu, r) }{\nu^{2n-2}}\,;
\qquad B_n (\nu, r)  = \frac1{\pi }
\int_0^\infty dy\, q_n(y )\,\,  G\left(\frac{\nu y}{2\kappa r} \right)
\end{equation}
where $G(x)$ is an even function of $x$:
\begin{equation} 
\label{eq:G}
G (x)=
\left\{
\begin{array}{ll}
\frac{\theta(1-x^2)}{\sqrt{1-x^2}}
& \quad\mbox { for } \alpha =0 \\
 \theta(1-x^2) \left\{
 1-4x^2+\frac{
 3 x (4x^2-3)
 }{\sqrt{1-x^2}}\sin^{-1} x
 \right\}
 + \theta(x^2-1)
 \left\{
 \frac{1}{\sqrt{x^2-1}}
 \frac{1- 3\log \left(\sqrt{x^2-1}+x\right)}{
 \left(\sqrt{x^2-1}+x\right)^3}\right\}
   & \quad\mbox { for } \alpha =3/2
    \end{array}
    \right.\,.
\end{equation}

The $\kappa \to \infty$ limit of the Casimir energy 
is given as 
$E_c^{(0)}(a) = {0.002819 }/{a}$,
the commutative result~\cite{{sphere-comm}}
since  $ G(0)=1$.
Its correction up to $O(1/\kappa^2)$
is given~\cite{KRY-asym} in terms of
$ \Delta {\cal E}_1 (r) $, 
$ \Delta {\cal E}_2 (r) $
and
\begin{align}
\label{rest}
\sum_{n \ge 3} \Delta  {\cal E}_n (a)
&=\frac {G_1}{\pi a}
\sum_{n \ge 3}\sum_{l} \frac {1 } {\nu^{2n-2}}
\int_0^{1/b}  dy \, q_n(y) (by)^2  + O(1/\kappa^3)
\end{align}
where
$ \Delta {\cal E}_n (r) 
= {\cal E}_n (r) - {\cal E}_n^{(0)} (r)$  
with ${\cal E}_n^{(0)} (r)
={\rm limit}_{\kappa \to \infty} {\cal E}_n (r)\,$,
$b =\frac{\nu }{2\kappa a}$ and
$G_1 =\left. \frac12 \frac{d^2}{dx^2} G(x) \right|_{x=0}\,$.
%%%%%%%%%%%%%%%%%%%%%%%%%%%%%%%%%%%%%%%%%%%%%%%
\section{Casimir energy when $\alpha=0$ and $\alpha=3/2$}
\label{sec:3}
%%%%%%%%%%%%%%%%%%%%%%%%%%%%%%%%%%%%%%%%%%%%%%%

When one neglects the measure factor ($\alpha=0$), 
one has $
B_1(\nu,a) \equiv B_1(b) 
= - { (1+16 b^2)}/{(128(1+b^2)^{5/2})}$
and 
$B_2(\nu,a)\equiv B_2(b) 
= {(35+174 b^2-7808 b^4+23552 b^6-4094 b^8)}/
 {(32768  (1+b^2)^{11/2})} \,$. 
${\cal E}_1(r)$ can be evaluated as
a series expansion in $1/(\kappa a)$ 
using a formula in~(\ref{approx:sum-integ})
\begin{equation} \label{E1}
{\cal E}_1(r)
= \frac1r \sum_{l=0}^\infty {B_1(\nu,r)}
= \frac1r \left(
    2\kappa r \int_{0}^\infty db\,B_1(b)
    +O(\kappa r)^{-3} \right)
=- \frac{3 \kappa} {32} +\frac{O(\kappa r)^{-3}}r \,.
\end{equation}
Thus  ${\cal E}_1(a) - {\cal E}_1(\eta R)$
is the order of $O(\kappa a)^{-3}$. 
To evaluate ${\cal E}_2(r)$,
one needs to take care of the fictitious 
singularity at $b=0$ using~(\ref{approx:sum-integ}):
One may put,
$B_2(b) = B_2^{\rm ref}(b) + (  B_2(b) - B_2^{\rm ref}(b))$
with $ B_{2}^{\rm ref}(b)= \frac{1}{32768}\left(\frac{35}{1+b^2}\right)$
($B_2^{\rm ref}(0) =B_2(0)$
and $B_2^{\rm ref}({b\to \infty} )=0 $)
to sum exactly $B_2^{\rm ref}(b) $, and 
use~(\ref{approx:sum-integ}) for the rest of the sum:
\begin{equation}\label{E2}
{\cal E}_2 (r)
= \frac{35 \pi^2}{65536r}
+\frac{35\pi }{131072 \kappa r^2}
+ \frac{O(\kappa r)^{-3}}r \,.
\end{equation} 
(One can check the result does not depend on
the explicit choice of  $ B_{2}^{\rm ref}(b)$).
The $\kappa$ independent term is contained in $E_c^{(0)}(a)$ 
and  the rest gives the deformed correction
$\Delta {\cal E}_2 (r)$. Finally,  $G_1 = 1/2$ in~(\ref{rest}) yields
\begin{equation}\label{E3}
\sum_{n \ge 3} \Delta  {\cal E}_n (a)
= \frac{J_1} {8\pi a(\kappa a)^2}  + O(1/\kappa^3)
\end{equation}
where $J_1\simeq 0.001713$ is evaluated in  Ref.~\cite{KRY-asym}. 
Combining Eqs.~(\ref{E1}), (\ref{E2}), and (\ref{E3}), 
one has the Casimir energy,
\begin{eqnarray}
\label{casimir-without-measure}
E_{c} = \frac{1}{a}\left(0.002819
    +\frac{35\pi }{131072 \kappa a}
   +\frac{0.00006816}{(\kappa a)^{2}}
   +O\left(\frac 1{\kappa a}\right)^3\right)\,.
\end{eqnarray}

%%%%%%%%%%%%%%%%%%%%%%%%%%%%%%%%%%%%%%%%
%{Casimir energy when $\alpha=3/2$}
%%%%%%%%%%%%%%%%%%%%%%%%%%%%%%%%%%%%%%%%
When the measure factor is included ($\alpha=3/2$),
one may conveniently arrange $B_n(\nu,r)$ into 3 pieces
$B_n(\nu,r)\equiv  (I_n(b) + K_n(b) +R_n(b))/\pi $
where 
\begin{align}
I_n(b) &\equiv
\int_0^{1/b}dy\, (1-4b^2 y^2)\,  q_{n}(y) %\label{I} 
\,,\quad
R_n(b)\equiv
\int_{1/b}^{\infty}dy \,\, q_n(y) G(b y)
\nn %\label{R}\,.
\\
K_n(b) &\equiv \int_0^{1/b}dy\,
\left( G(b y)- (1-4b^2 y^2) \right) q_{n}(y)\,.
%\left\{\frac{ 3 by (4b^2y^2-3)
% }{\sqrt{1-b^2y^2}}\sin^{-1}(by)
% \right\}  
%\label{K}
\nn
\end{align}
Explicitly,
$I_1(b)= - \frac1{64}\left({b \left(108 b^4+179
   b^2+41\right)}/{\left(b^2+1\right)^2}+\left(1-108
   b^2\right) \cot ^{-1}(b)\right)$
is $O(b^{-3})$ for large $b$
and is finite for $b\rightarrow 0$.   
Using~(\ref{approx:sum-integ}) one has
$\sum_{l=0}^\infty I_1(b) 
= {\kappa a}/{16} - {5}/{(384\kappa a)}
+O (1/{(\kappa a)^3} ) \,$.
$K_1(b)$ is easy to calculate
if the sum over $l$ is done first.
Note that $\sum_{l=0}^\infty {q_1(\frac{s}{b})}/{b}
= \frac 18 ( -4s^2 \,S_2(s)+ 5s^4 \,S_3(s))$
where 
\begin{equation} \label{summ}
S_n(s)\equiv \sum_{l=0}^\infty \frac{b}{(b^2+s^2)^n}
    =\frac{\kappa a}{(n-1)s^{2(n-1)}}+
        \frac{1}{48\kappa as^{2n}}
        +O\left(\frac 1{\kappa a} \right)^3 \,.
\end{equation}
Thus, summing $K_1(b)$ over  $l$ 
(change of the integration variable $y \to y/b$ is used) 
gives 
$\sum_{l=0}^\infty K_1(b) =
-{\kappa r}/{16} -  {(3C -2)}/{(64 \kappa r)}
+O(1/(\kappa r)^3)$
where $C\simeq 0.915966$ is the Catalan's constant.
The sum of $R_1$ over $l$ is done in the same way:
$\sum_{l=0}^\infty R_1(b)= -{(7 - 18 C +3 \pi)}/{(384\kappa a)}
+O\left(\frac 1{\kappa a} \right)^3$. 
Combining all the contributions one has 
\begin{equation}\label{E1:2}
\Delta \Big(\mathcal{E}_1(a)-\mathcal{E}_1(\eta R )\Big)
= -\frac{1}{a}\left(
    \frac{1}{128\,\kappa a}
    +O\left(\frac 1{ (\kappa a)^3} \right)\right)\,.
\end{equation}

Similarly, the rest of the terms are given as 
\begin{align}
 \label{E2:2}
&\Delta \Big(\mathcal{E}_2(a)-\mathcal{E}_2(\eta R )\Big)
 =\frac{1}{ a }\left(
    \frac{-191+108 \pi}{9216 \pi(\kappa a)}
    +O\left(\frac 1{(\kappa a)^3} \right)   \right)
    \\
&\sum_{n \ge 3} \Delta  {\cal E}_n (a) 
= \frac1a\left(\frac{G_1 J_1}{4\pi (\kappa a )^2} + +O\left(\frac 1{(\kappa a)^3} \right)  \right)
\end{align}
with $G_1 =-13$ and $J_1\simeq 0.001713$.
Thus, the Casimir energy is given as 
\begin{eqnarray}
\label{casimir-with-measure}
E_c = \frac{1}{a}\left(0.002819-\frac{0.002691}{\kappa a}
-  \frac{0.001772}{(\kappa a)^2}  
+O\left(\frac 1{(\kappa a)^3} \right)\right)\,.
\end{eqnarray}

%%%%%%%%%%%%%%%%%%%%%%%%%%%%%%%%%%%%%%%%%%%%%
\section{Summary and Discussions } \label{sec:4}
%%%%%%%%%%%%%%%%%%%%%%%%%%%%%%%%%%%%%%%%%%%%%%%

The Casimir energy of massless scalar field 
in $\kappa$-Minkowski space-time 
is presented in (\ref{casimir-without-measure})
when the measure factor is neglected,
and in (\ref{casimir-with-measure}) 
when the measure factor is taken care of. 
The negative mode (anti-particle) contribution 
is formally given if $\kappa \to - \kappa$
in (\ref{casimir-without-measure}) and 
(\ref{casimir-with-measure}).  
However, the formula obtained in (\ref{casimir-without-measure})
and (\ref{casimir-with-measure}) 
is valid only for $\kappa > 0$.
One can extend the result to $\kappa<0$
using the even property of $G(x)$ under the sign change 
of $\kappa$ to $-\kappa$:
\begin{align}
E_{c}^{\alpha=0}(a) 
&= \frac{1}{a}\left(0.002819
    +\frac{0.0008389}{|\kappa a|}
   +\frac{0.00006816}{(\kappa a)^{2}}
   +O\left(\frac 1{\kappa a}\right)^3\right)\,,
\nn\\
E_{c}^{\alpha=3/2}(a) 
&= \frac{1}{a}\left(0.002819-\frac{0.002691}{|\kappa a|}
-  \frac{0.001772}{(\kappa a)^2}+O\left(\frac{1}{\kappa a}\right)^{3}\right) 
\label{pa-symmetry}\,.
\end{align} 
This demonstrates that the particle contribution 
and the antiparticle contribution are equal. 
It should be emphasized that 
the symmetric result (\ref{pa-symmetry})
originates from the anlytic property  
of the dispersion relation (\ref{omega})
(the integration range in (\ref{En-analytic}) is from 0 to $\infty$). 
This is contrasted with 
the asymmetric ordering case 
in~\cite{KRY-asym} where 
the presence of a branch-cut 
spoils the particle and anti-particle symmetry.

The dispersion relation (\ref{omega}) is obtained 
from the Casimir invariant 
$ M_s^2(p)
=\left(2 \kappa \sinh {p_0}/
{(2 \kappa )}\right)^2 - {\bf p}^2 = 0 \,$. 
When $ |{\bf p}| \ge 2 \kappa$,
however, there appears another real mode,
so called the high momentum mode, 
from the relation 
$M_s^2(p)
=\left(2 \kappa \sinh 
{p_0}/{(2 \kappa) }\right)^2
    - {\bf p}^2 = -4 \kappa^2\,$.
Considering the result of black body 
radiation~\cite{KRY-field},
where the high momentum mode spoils the 
$\kappa \to \infty$ limit,
one needs to exclude the high momentum mode 
on the mass-shell condition. 

Finally, it is worth to mention that 
the $\kappa$ correction to the 
Casimir energy (\ref{pa-symmetry}) is of $O(1/\kappa)$.
This is contrasted with results shown 
in the parallel plate case.
According to~\cite{cougo} %and~\cite{nam}
the Casimir energy of electromagnetic field
(electric and magnetic modes) per unit area
is of the $O(1/\kappa^2)$.

%%%%%%%%%%%%%%%%%%%%%%%%%%%%%%%%%%%%%%%%%%%%%%%%
\begin{acknowledgments}
This work was supported in part 
by the Korea Research Foundation Grant funded by
Korea Government 
(MOEHRD, Basic Research Promotion Fund, (KRF-2005-075-C00009; H.-C.K.)
and(KRF-2007-2-313-C00153:R) and  
by the the Center for Quantum Spacetime (CQUeST)
of Sogang University with grant (R11-2005-021).
It is also acknowledged by R that 
this report has been completed during his visit to 
Korea Institute for Advanced Study (KIAS).
\end{acknowledgments} 

\begin{appendix}

%%%%%%%%%%%%%%%%%%%%%%%%%%%%%%%%%%%%%%%%%%%%%%%%
\section{Approximation of summation by using integration}
\label{appendix}
%%%%%%%%%%%%%%%%%%%%%%%%%%%%%%%%%%%%%%%%%%%%%%%%
Let us consider the following summation:
$\sum_{l=0}^\infty f(\frac{l+1/2}{L}) \frac{1}{L}\,$
where  $\lim_{x\rightarrow \infty}f(x)=0$.
For $L \rightarrow \infty$, this becomes the integral $\int_0^\infty f(x) dx$ for non-singular continuous function $f(x)$. However, for large but finite $L$, this is not the case, and we are presenting 
a reasonable approximation of this sum as an integral.
Using the Taylor series at $x_c=(l+1/2)/{L}$, $f(x) = \sum_{n=0}^\infty {f^{(n)}(x_c)} (x-x_c)^n/{n!}\,$~  one can put 
\begin{align} 
\int_{x_c-1/(2L)}^{x_c+1/(2L)} f(x) dx \
&=\frac{f(x_c)}{L} 
+ \sum_{k=1}^\infty f^{(2k)}(x_c) \frac{2}{ (2L)^{2k+1}(2k+1)!} \,.
\end{align}
Successive application of this formula gives 
\begin{equation}
f(x_c)
=L\int_{x_c-1/(2L)}^{x_c+1/(2L)} \left[f(x) -\frac{1}{24L^2} f^{(2)}(x)\right] + \sum_{k=2}^\infty  \left(\frac{2k(2k+1)}{6}-1\right)\frac{f^{(2k)}(x_c)}{ (2L)^{2k}(2k+1)!}\,.
\end{equation}
Noting that the last term is of $O(L^{-4})$, one has 
\begin{equation} 
\label{approx:sum-integ}
\sum_{l=0}^\infty f(\frac{l+1/2}{L}) 
   =L\int_0^\infty dx\,f(x) +\frac{f'(0)}{24L} +O(L^{-3}) \,.
\end{equation}
For the special case when $f(x)$ is an even function, 
the summation over $f(x)$
is approximated by its integration only
since the right-hand side of (\ref{approx:sum-integ}) 
does not contain any even number of derivatives to all orders in $1/L$.  
For example, if $f(x)=(1+x^2)^{-1}$, (\ref{approx:sum-integ}) gives
$L\int_0^\infty dx\,f(x)= L\pi /2$ to all orders in $1/L$.
On the other hand, one may perform exact summation 
to have $(L \pi/2) \tanh(L\pi)\, $.
The difference of the two is of $O(e^{-L \pi})$, which is exponentially small for large $L$. This is just an example of the fact that the exponentially small correction is not well represented by a Taylor series.

\end{appendix}

%%%%%%%%%%%%%%%%%%%%%%%%%%%%%%%%%%%%%%%%%%%%%%%%

\end{document}